\documentclass[sigconf]{acmart}

\AtBeginDocument{%
  \providecommand\BibTeX{{%
    \normalfont B\kern-0.5em{\scshape i\kern-0.25em b}\kern-0.8em\TeX}}}

\copyrightyear{2023}
\acmYear{2023}
\setcopyright{acmlicensed}\acmConference[WWW '23]{Proceedings of the ACM
Web Conference 2023}{May 1--5, 2023}{Austin, TX, USA}
\acmBooktitle{Proceedings of the ACM Web Conference 2023 (WWW '23), May
1--5, 2023, Austin, TX, USA}
\acmPrice{15.00}
\acmDOI{10.1145/3543507.3583244}
\acmISBN{978-1-4503-9416-1/23/04}

\usepackage{bm}
\usepackage{algorithm}
\usepackage{algorithmic}
\usepackage{multicol}
\usepackage{multirow}
\usepackage{balance}

\begin{document}

\title{Exploration and Regularization of the Latent Action Space in Recommendation}


\author{Shuchang Liu}
\affiliation{%
  \institution{Kuaishou Technology}
  \city{Beijing}
  \country{China}}
\email{liushuchang@kuaishou.com}

\author{Qingpeng Cai}
\affiliation{%
  \institution{Kuaishou Technology}
  \city{Beijing}
  \country{China}}
\email{caiqingpeng@@kuaishou.com}

\author{Bowen Sun}
\affiliation{%
  \institution{Peking University}
  \city{Beijing}
  \country{China}}
\email{bwzdbml@gmail.com}

\author{Yuhao Wang}
\affiliation{%
  \institution{City University of Hong Kong}
  \city{Hong Kong}
  \country{China}}
\email{yhwang25-c@my.cityu.edu.hk}

\author{Ji Jiang}
\affiliation{%
  \institution{Peking University}
  \city{Beijing}
  \country{China}}
\email{johnjim0816@gmail.com}

\author{Dong Zheng}
\affiliation{%
  \institution{Kuaishou Technology}
  \city{Beijing}
  \country{China}}
\email{zhengdong@kuaishou.com}

\author{Kun Gai}
\affiliation{%
  \institution{Unaffiliated}
  \city{Beijing}
  \country{China}}
\email{gai.kun@qq.com}

\author{Peng Jiang$^\dagger$}
\thanks{$\dagger$ \text{Corresponding author}}
\affiliation{%
  \institution{Kuaishou Technology}
  \city{Beijing}
  \country{China}}
\email{jiangpeng@kuaishou.com}

\author{Xiangyu Zhao}
\affiliation{%
  \institution{City University of Hong Kong}
  \city{Hong Kong}
  \country{China}}
\email{xianzhao@cityu.edu.hk}

\author{Yongfeng Zhang}
\affiliation{%
  \institution{Rutgers University}
  \city{New Brunswick, NJ}
  \country{USA}}
\email{yongfeng.zhang@rutgers.edu}





\renewcommand{\shortauthors}{Liu and Cai, et al.}

\begin{abstract}
In recommender systems, reinforcement learning solutions have effectively boosted recommendation performance because of their ability to capture long-term user-system interaction.
However, the action space of the recommendation policy is a list of items, which could be extremely large with a dynamic candidate item pool.
To overcome this challenge, we propose a hyper-actor and critic learning framework where the policy decomposes the item list generation process into a hyper-action inference step and an effect-action selection step.
The first step maps the given state space into a vectorized hyper-action space, and the second step selects the item list based on the hyper-action.
In order to regulate the discrepancy between the two action spaces, we design an alignment module along with a kernel mapping function for items to ensure inference accuracy and include a supervision module to stabilize the learning process.
We build simulated environments on public datasets and empirically show that our framework is superior in recommendation compared to standard RL baselines.
\end{abstract}



\maketitle

\section{Introduction}\label{sec: intro}

Recommender Systems (RS) serve as one of the fundamental components for a wide range of web services including e-commerce, social media, news, and advertising. 
In recent years, studies have shown that the long-term interactions between users and the RS formulate a Markov Decision Process (MDP), where Reinforcement Learning (RL) methods can be used to further improve the predictive performances~\cite{afsar2021reinforcement} compared to traditional learning-to-rank solutions~\cite{liu2009learning}.
Rather than optimizing the immediate user response, the key insight behind RL is to maximize the cumulative reward over all the interactions over time.

When adopting this formulation in practice, the recommendation problem distinguishes itself from other RL tasks by the challenge of \textit{large dynamic discrete} action space.
For example, in an e-commerce system or video recommendation system, it may need to select for each request a recommendation list from a pool of millions of products, and the candidate pool grows every day.
This means that it would be unrealistic for tabular-based methods (e.g. Q-Learning, SARSA, Policy Iteration) that favor small action spaces and methods designed for fixed action spaces.
Though efforts have been made to alleviate this issue by decomposition of the recommendation list~\cite{ie2019slateq} into item-wise sub actions, learning a policy gradient over the entire candidate item pool is still an challenging task.

\begin{figure}[t]
    \centering
    \includegraphics[width=0.9\linewidth]{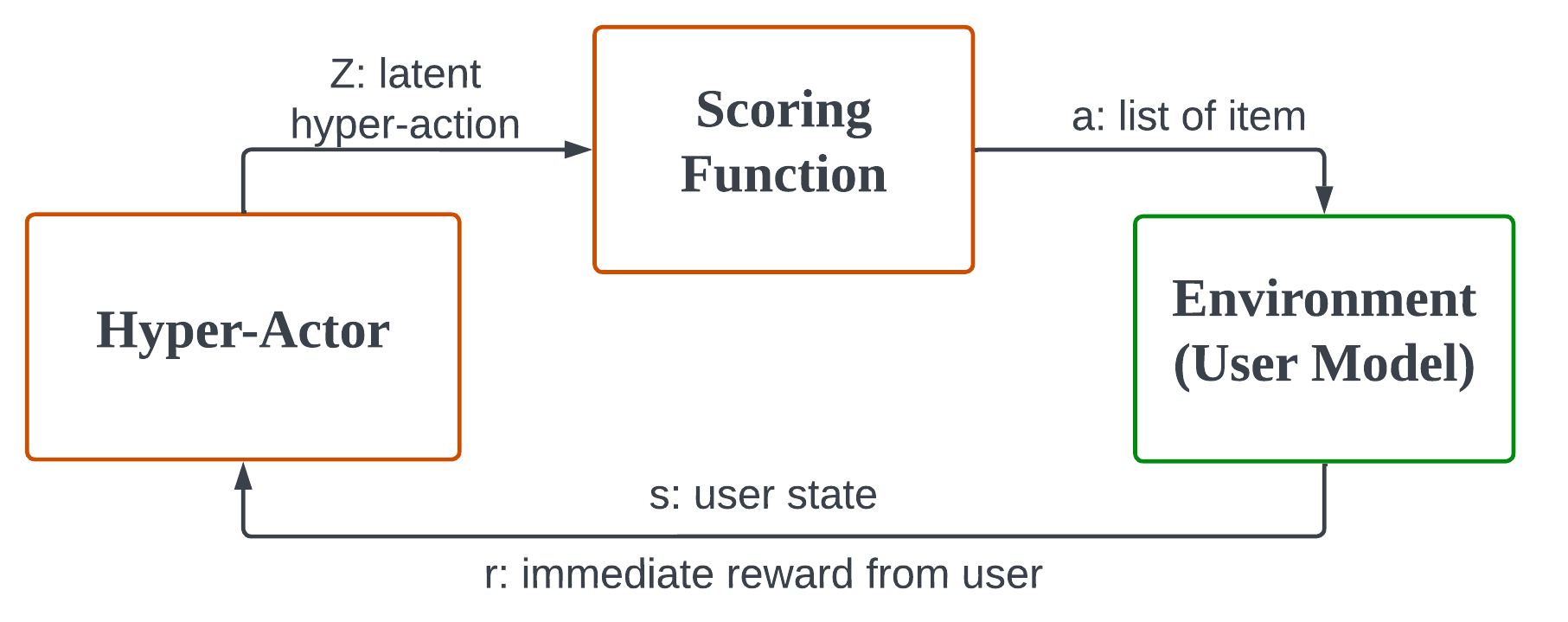}
    \caption{MDP formulation with latent hyper-action}
    \label{fig: mdp}
    \vspace{-10pt}
\end{figure}

Fortunately, this challenge has already been solved in early-age non-RL models like latent factor models~\cite{koren2009matrix} and two-tower collaborative filtering method~\cite{yi2019sampling}.
They learn a common latent space that can represent both the user requests and items (or item lists), so that the learned latent space can accommodate arbitrary number of items and is agnostic to the dynamic changes of the item pool.
In this work, we combine this insight into the RL-methods and focus on the list recommendation problem.
The latent representations of recommendation lists are generalized as hyper-actions as shown in Figure \ref{fig: mdp}.
In a forward inference, the policy first propose a vectorized \textit{hyper-action}, then this latent vector will serve as a deterministic function that rank, and finally select a list of items, denoted as \textit{effect-action}, from the candidate pool.

Note that this extra implicit inference step also induces new challenges to the RL solution:
Most importantly, it introduces inconsistency between the two actions spaces.
Specifically, we want to apply efficient end-to-end training on the hyper-actions but it is the effect-actions that actually interact with the users.
On the other hand, the most accurate latent representation of the discrete effect-action may not be exactly the same as that of the proposed hyper-action that selects it, so the inference accuracy is not guaranteed.
Additionally, it also introduces an extra exploration stage and it is unknown whether one should explore hyper-actions on the latent space or explore the discrete effect-action space.
All of these add up to the instability and uncertainty of the learning, so we need a solution that can regulate the two action spaces and stabilize the RL process.

To solve the aforementioned challenges, we propose a general Hyper-Actor Critic (HAC) learning framework that contains four components: the user state and hyper-action generator; the scoring function that maps hyper-actions into effect-actions (i.e. recommendation lists); the critic network that evaluates both the hyper-action space and the effect-action space; and an inverse mapping module that infers the hyper-action back based on the effect action.
During training, the backbone actor-critic learning paradigm is augmented with an alignment module that ensures consistency between two action spaces and a supervision module that improves the stability and effectiveness of RL.
The resulting framework generalizes many existing solutions to recommendation tasks like DDPG and Online/Offline Supervise Learning (SL).

We summarize our contribution as follows:
\begin{itemize}
    \item We propose a practical and efficient RL framework that learns to recommend a list of items from a large item pool through a latent hyper-action.
    \item We build online simulators based on public datasets and empirically show that the resulting framework achieves better performance compared to standard RL and SL solutions.
    \item We also point out that providing supervision and regulating the consistency between the hyper-action space and the effect-action space are helpful for improving the sampling efficiency and inference accuracy.
\end{itemize}

\section{Method}\label{sec: method}

\subsection{Problem Formulation}\label{sec: problem_formulation}

Here we consider the session-based recommendation scenario where the system considers an item pool of size $N$, denoted as $\mathcal{I}$, and for each user we observe the static user features $\bm{u}$ and the interaction history of a session $x_{1:t} = \{x_1,\dots,x_t\}$ where $x_i\in\mathcal{I}(1\leq i\leq t)$.
And the goal is to interactively recommend lists of items to users that maximizes user cumulative reward over the session of interactions.
As we have described in section \ref{sec: intro}, we emphasize the existence of the latent action space and formulate the recommendation task as a modified Markov Decision Process (MDP) as captured in Figure \ref{fig: mdp}.
Then, the MDP components with the latent action space become:
\begin{itemize}
    \item $\mathcal{S}$: the continuous representation space of user state.
    \item $\mathcal{A}$: the final effect-action space corresponds to the possible recommendation lists. Without loss of generality, we consider the list of fixed size $k$ so the action space is $\mathcal{A}=\mathcal{I}^k$.
    \item $\mathcal{Z}$: the latent hyper-action space which is a continuous vector space that encodes how the effect-action will be selected. We assume a many-to-one mapping function $f: \mathcal{Z} \rightarrow \mathcal{A}$.
    \item $\mathcal{R}$: The cumulative reward function that estimates the user feedback in the user session, and the immediate reward $r(s,a)$ captures the single step reward when taking an action $a\in\mathcal{A }$ on state $s\in\mathcal{S}$.
\end{itemize}
Note that there is an implicit transition model $\mathcal{P}(s_t^\prime|s_t,a_t)$ that describes the probability of reaching a certain new state $s_t^\prime$ after taking action $a_t$ on state $s_t$. 
In RS, this transition function is integrated into the user state encoder and is usually modeled by a sequential model that takes the user interaction sequence as input.

At each interaction step in a user session, given a user's current state $s_t\in\mathcal{S}$ (e.g. portrait and history interactions), the recommendation policy $\pi_\theta(a_t|s_t)$ first infer a hyper-action representation $Z_t$ and then generates a list of items as the effect action $a_t$.
The user's feedback along with the updated user state is returned from the user environment and the reward function $r(s_t, a_t)$ is assumed as given so is regarded as part of the environment.
The goal is to find an optimal recommendation policy $\pi^*(a_t|s_t): \mathcal{S}\rightarrow\mathcal{A}$ that maximizes the expected cumulative reward throughout the session:
\begin{equation}
    \mathbb{E}_{\tau\sim\pi}[\mathcal{R}(\tau)]=\mathbb{E}_{\tau\sim\pi}\Big[\sum_{t=0}^{|\tau|}\gamma^t r(s_t,a_t)\Big]
\end{equation}
where $\tau = [(s_0,a_0,r_0),(s_1,a_1,r_1),\dots]$ denotes the sampled trajectories, and $\gamma\in[0,1)$ denotes the discount factor for future reward.

\subsection{Overall Framework}

\begin{figure*}[t]
    \centering
    \includegraphics[width=0.9\textwidth]{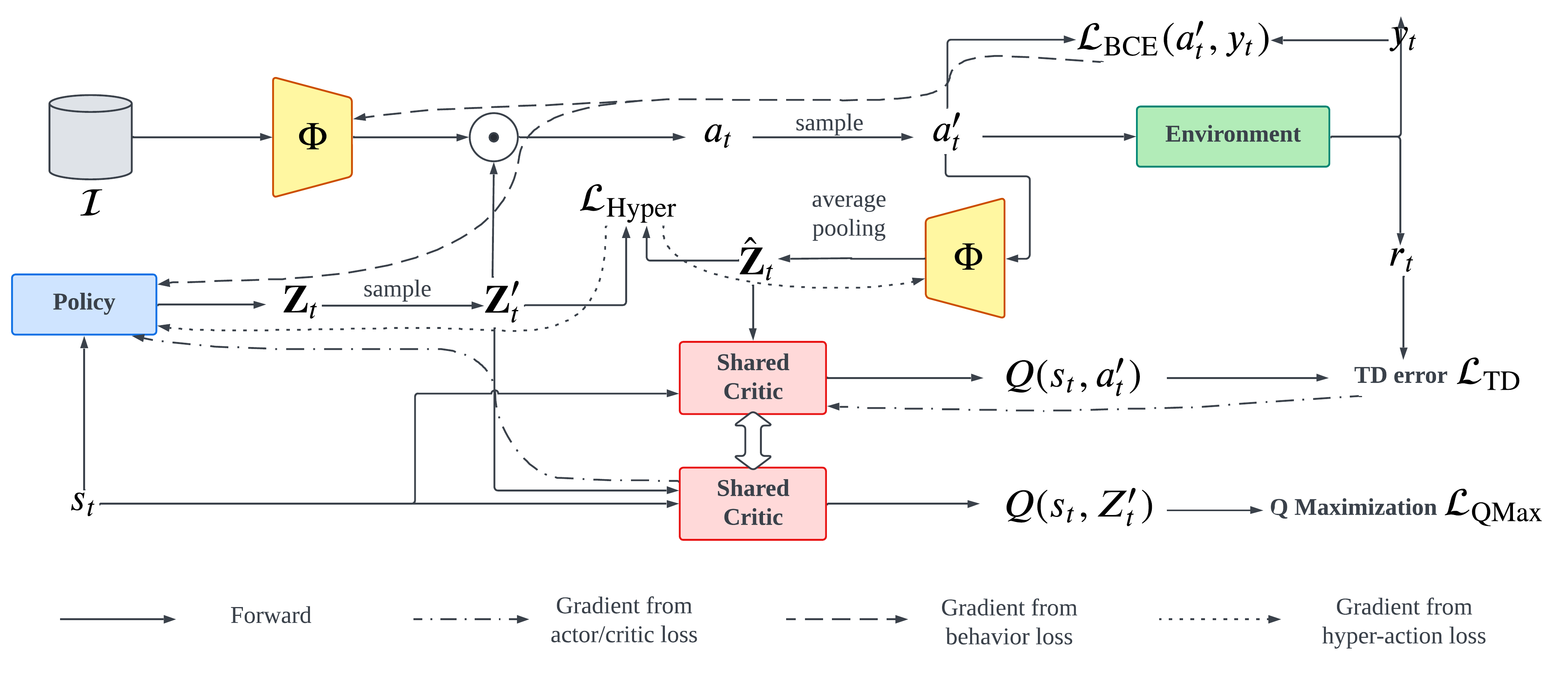}
    \caption{Hyper-Actor Critic (HAC) learning framework. $\odot$ represents the scoring function that selects items from $\mathcal{I}$, }
    \label{fig: solution_framework}
\end{figure*}

We present our framework as Figure \ref{fig: solution_framework}, and denote it as the Hyper-Actor Critic (HAC) learning method.
The recommendation policy $\pi(a_t|s_t)$ is decomposed into an hyper-actor network $P(Z_t|s_t)$ that generates a vectorized hyper-action and a ranking scorer $P(a_t|Z_t)$ that select the final recommendation list based on the hyper-action.
Then we propose to share the critic network between action spaces so that it can evaluate either the hyper-action or the final effect-action (i.e. the recommendation list).
Our framework uses DDPG~\cite{timothy2016ddpg} as foundation, but differently, we address the importance of using different action spaces for actor learning and critic learning.
Specifically, we optimize the critic based on effect-actions to guarantee the accuracy of action/state evaluation, and use hyper-actions to optimize the actor so that efficient end-to-end training and exploration can be applied.
To ensure consistency between the two different action spaces, we also learn an inverse pooling module with item kernel functions to infer the hyper-action back from the effect-action.
This means that the evaluation of the two action spaces will share the same critic, and the knowledge learned from the effect-action can be transferred to hyper-actions.
To stabilize the learning process, we also include supervision on the effect-action using immediate user responses.

\subsection{User State and Hyper-Actor}\label{sec: method_state_encoder}

In the case of RS, the observable information of a user usually includes the up-to-date user interaction history $\bm{x}_{1:t}$ and the static demographic features $\bm{u}$.
Then one can use any sequential model to encode this information and infer the dynamic user state:
\begin{equation}
    \bm{s}_t = \mathrm{StateEnc}(\Phi(x_1),\dots,\Phi(x_t),\bm{u})
\end{equation}
where the item kernel function $\Phi$ maps the items' raw features into a dense embedding in the kernel space.
We also use a user kernel function to map the user features $\bm{u}$ into the same kernel space of items, then concatenate the sequence of history items with the user embedding as the new sequence.

To encode user features and histories, there exist several feasible sequential models such as~\cite{jannach2017gru4rec,kang2018sasrec,sun2019bert4rec}, and we use the state encoder in SASRec~\cite{kang2018sasrec} as our backbone since it can capture both the sequential patterns in the dynamic history and the user-item correlations through the self-attention mechanism.
With the state encoded, a vectorized representation (i.e. the hyper-action) is inferred by an Multi-Layer Perceptron (MLP) module:
\begin{equation}
    \bm{Z}_t = \mathrm{MLP}(\bm{s}_t)
\end{equation}
We assume that the distribution of this hyper-action follows the standard Gaussian $\mathcal{N}(\bm{Z}_t, \sigma_Z^2)$ and we can use the reparameterization trick to engage the end-to-end training.

\subsection{Scoring Functions and Effect Action}\label{sec: method_scorer} 

Given $\bm{Z}_t$ that contains sufficient information of user preferences, we can assume that the selection of items only depends on this latent vector.
In other words, we have conditional independence:
\begin{equation}
    P(a_t|\bm{s}_t,\bm{Z}_t) = P(a_t|\bm{Z}_t)
\end{equation}
Note that this setting also indicates that the inferred $\bm{Z}_t$ is a \textit{hyper-action} that can be considered as the parameters of the later item selection module.
As a result, the overall recommendation policy follows $\pi(a_t|s_t) \sim P(a_t|\bm{Z}_t)$. 

When generating the effect-action, we select items from the candidate pool according to a scoring function parameterized by $\bm{Z}_t$. 
The scoring function provides a ranking score for each of the candidate items in $\mathcal{I}$, and the final recommendation list is generated through either top-$k$ selection or categorical sampling.
Taking the SASRec model as an example, the scoring function is a dot product between the kernel item embedding and the encoded user state:
\begin{equation}
    \mathrm{score}(i|Z_t) = \Phi(i)^\top \bm{Z}_t
\end{equation}
Note that the parameters of the item kernel function $\Phi$ are not considered as part of the hyper-action $\bm{Z}_t$ since it is independent of the given user state.
In order to engage efficient learning and inference, one can assume that selection of each item is conditionally independent of each other:
$P(a_t|Z_t) = \prod_{i\in a_t} P(i|Z_t)$, similar to the slate decomposition in~\cite{ie2019slateq}.
Then, we can define the selection or sampling probability of an item as $P(i|Z_t) = \mathrm{softmax}_\mathcal{I}(\mathrm{score}(i|Z_t))$.

\subsection{Shared Critic and The Inverse Module}\label{sec: method_shared_critic}

The purpose of the critic is to accurately evaluate the long-term quality of the state-action pair (e.g. Q function in DDPG) or the expected value of the given state (e.g. V function in A2C) so that it can effectively guide the actor learning and the action exploration.
Compared to the standard RL framework, the new problem setting allows us to evaluate either the hyper-action with $Q(s_t, \bm{Z}_t)$ or the effect-action with $Q(s_t, a_t$) or both.
In order to ensure consistent evaluation of the actions from different spaces, we propose to transfer knowledge between $Q(s_t,a_t)$ and $Q(s_t, Z_t)$ through a shared critic network.
As shown in Figure \ref{fig: solution_framework}, this shared critic is a mapping function $g: \mathcal{S}\times \mathcal{Z}\rightarrow \mathbb{R}$, that takes the user state $s_t$ and the action embedding in the kernel space $Z_t$.
Or equivalently, $Q(s_t,Z_t) = g(s_t,Z_t)$.
To evaluate the effect-action, an inverse module $h$ is introduced to infer the hyper-action back:
\begin{equation}
    \hat{\bm{Z}}_t = h(a_t) = \mathrm{pooling}(\Phi(i)|i\in a_t)
\end{equation}
and the evaluation becomes $Q(s_t,a_t) = g(s_t, \hat{\bm{Z}}_t)$.
In practice, we found that the average pooling of the item embedding in the kernel space generates the most stable result, though there are infinitely many latent $\bm{Z}$ that can generate the same list.
Compared to existing solutions that infer the latent action using adjacent states like PG-RA~\cite{chandak2019learning}, we believe that the effect-action in recommendation task has sufficient information to recover the hyper-action.
In addition, we use a reconstruction loss to further regulate the consistency between the hyper-action and the effect-action through an alignment loss function.
As we will describe in the next section, it ensures that the generated hyper-action $\bm{Z}_t$ is in the valid region close to the candidate items in the kernel space.

\subsection{Overall Learning Framework}\label{sec: method_learning}

The overall optimization process is a modified actor-critic learning framework that consists of a critic loss, an actor loss, a hyper-actor loss, and a supervised loss.
And a experience replay buffer $\mathcal{D}$ will collect the sample records in the form of $(s_t, a_t, r(s_t,a_t), s_{t+1}, d)$ where $d$ is the termination signal indicating whether the user has left.
The critic loss aims to train an accurate evaluator that captures the patterns in the quality of actions:
\begin{equation}
\begin{aligned}
    \mathcal{L}_\mathrm{TD} & = \mathbb{E}_{\mathcal{D}}\Big[(r(s_t, a_t) + \gamma (1-d)Q(s_{t+1},a_{t+1}) - Q(s_t,a_t))^2\Big]\label{eq: critic_loss}
\end{aligned}
\end{equation}
where $a_{t+1}$ is generated by the recommendation policy using greedy method (equivalent to finding $\arg\max Q(s_t, a_t)$ when the Q function is accurate).
This is a standard TD error and we only calculate Q for the effect-action when learning the critic to ensure the accuracy of evaluation.
Note that in DDPG-based methods, each actor and critic is paired with a target network which is used to estimate future state-action pairs $Q(s_{t+1}, a_{t+1})$.
These target networks adopt elastic updates from the actor and critic so that their slow optimization can stabilize the learning.

Then, with an accurate critic as the evaluator, we can efficiently learn our actor by maximizing the Q-value of the inferred action, which is equivalent to minimizing the following actor loss:
\begin{equation}
    \mathcal{L}_\mathrm{QMax} = \mathbb{E}_\mathcal{D}\Big[Q(\bm{s}_t, \bm{Z}_t)\Big]\label{eq: actor_loss}
\end{equation}
where $\bm{Z}_t$ is inferred by the hyper-actor as described in section \ref{sec: method_state_encoder}.
Note that the learning of critic and actor uses different action spaces, so we need to align the two spaces to avoid the mode collapse~\cite{srivastava2017veegan} of the generated hyper-action.
In our solution, we use the L2 regularizer to ensure this consistency:
\begin{equation}
    \mathcal{L}_\mathrm{Hyper} = \mathbb{E}_D\Big[\|\bm{Z}_t - \hat{\bm{Z}}_t\|^2\Big]\label{eq: hyper_actor_loss}
\end{equation}
where $\bm{Z}_t$ is produced by the hyper-action based on state $\bm{s}_t$, and $\hat{\bm{Z}}_t$ is generated by first greedily select the effect action using $\bm{Z}_t$ as described in section \ref{sec: method_scorer} and then reconstruct the hyper-action back using the inverse module as described in section \ref{sec: method_shared_critic}.

Additionally, to better stabilize the RL and exploit the detailed user response signals on each item, we also include a supervised learning objective based on the effect-action.
\begin{equation}
    \mathcal{L}_\mathrm{BCE} = \mathbb{E}\Big[\sum_{i\in a_t}y_{t,i}\log P(i|\bm{Z}_t) + (1-y_{t,i})\log (1-P(i|\bm{Z}_t))\Big]\label{eq: supervise_loss}
\end{equation}
which is a binary cross-entropy loss where $y_{t,i}$ is the ground truth user response on the exposed item $i$.
We remind readers that there are other advanced supervision and regularization methods that can accommodate RL-based models and potentially extend our framework.
For example, ~\cite{fujimoto2021minimalist} could supervise the effect-action space for off-policy training, and ~\cite{xin2022sac} is well-suited for the distance control on the hyper-action space.
As a summary, we present the resulting learning paradigm as algorithm \ref{alg: solution}.
And note that the parameters of the inverse module come from the item kernel, so both Eq.\eqref{eq: hyper_actor_loss} and Eq.\eqref{eq: supervise_loss} update them as in line 8-9.

\begin{algorithm}[t]
\caption{Hyper-Actor Critic Training}
\label{alg: solution}
\begin{algorithmic}[1]
\STATE \textbf{procedure} HAC
\STATE Initialize all trainable parameters in the actors, critics, and the item kernel function.
\STATE Initialize replay buffer $\mathcal{B}$.
\WHILE{Not Converged, in each iteration}
    \STATE Apply current policy in running episodes, collect and store samples to $\mathcal{B}$.
    \STATE Sample mini-batch of $(\bm{s}_t, a_t, r(\bm{s}_t,a_t), \bm{s}_{t+1}, d) \sim \mathcal{B}$.
    \STATE Update actor and critic with loss Eq.\eqref{eq: actor_loss} and Eq.\eqref{eq: critic_loss}.
    \STATE Update actor and the kernel with loss Eq.\eqref{eq: hyper_actor_loss}, if any action alignment.
    \STATE Update actor and the kernel with loss Eq.\eqref{eq: supervise_loss}, if any supervision.
\ENDWHILE
\end{algorithmic}
\end{algorithm}


\subsection{Exploration in Hyper-Action Space and Effect-Action Space}\label{sec: method_exploration}

Readers may notice that the inclusion of latent hyper-action also introduces an extra action sampling step as shown in Figure \ref{fig: solution_framework}, so the resulting framework allows both the sampling on the hyper-actions space (e.g. by adding Gaussian noise) and the sampling on the effect-action space (e.g. categorical sampling of items based on ranking scores).
Theoretically, this indicates that the sampling probability of effect-actions should be described as the following:
\begin{equation}
    P(a_t|\bm{s}_t) = \int_{Z_t} P(a_t|\bm{Z}_t) P(\bm{Z}_t|\bm{s}_t)
\end{equation}
When the effect-action generation is deterministic, the exploration only depends on the sampling of $\bm{Z}_t$, similar to that in~\cite{chandak2019learning}; 
and if the hyper-actor is deterministic, the exploration only depends on the effect-action sampling as in standard policy gradient methods.
Note that the variance $\sigma_Z^2$ of the hyper-action controls the uncertainty of the inferred latent action embedding, and it is critical to find an adequate value that can improve the exploration effectiveness in RL.
On one hand, giving a variance that is too small will limit the exploration of new actions resulting in sub-optimal results; On the other hand, making the variance too large will induce unstable action exploration that hardly converges.

\begin{figure}[t]
    \centering
    \includegraphics[width=\linewidth]{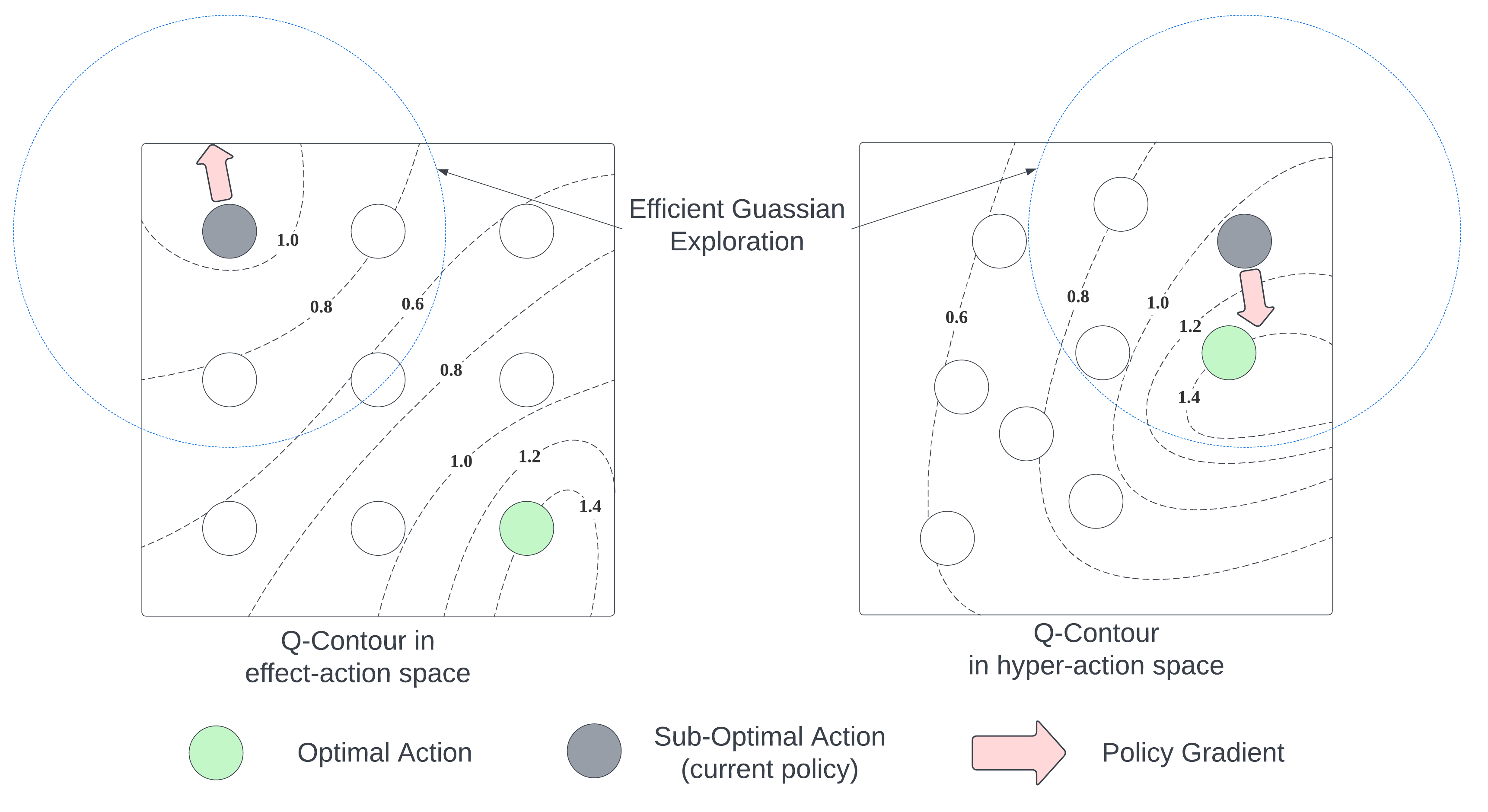}
    \caption{Exploration in Different Action Spaces}
    \label{fig: exploration}
\end{figure}

In general, we would like to take advantage of the efficient learning and exploration of the hyper-action space, so it becomes critical to align the distribution of $\bm{Z}_t$ and the embedded item in the kernel space, as we mentioned in section \ref{sec: method_learning}.
As we showcase in Figure \ref{fig: exploration}, the item kernel function helps increase the expressiveness of the policy by folding the action space where exploration could be more efficient.
Though we are skeptical whether there is a guarantee for all RL solutions that explores the latent action space, we will empirically show the effectiveness of encoding both users and items into the same kernel space and regulating the action with the inverse pooling module using Eq.\eqref{eq: hyper_actor_loss} in section \ref{sec: experiment_ablation}.

\section{Experiments}\label{sec: experiments}

\subsection{Experimental Settings}\label{sec: experiment_setup}

\subsubsection{Datasets:} We include three public datasets in our experiments: \textbf{RL4RS}\footnote{https://github.com/fuxiAIlab/RL4RS} is a session-based dataset~\cite{2021RL4RS} that is first introduced in the BigData Cup 2021 to boost RL research; 
\textbf{ML1M}\footnote{https://grouplens.org/datasets/movielens/1m/} is the MovieLens data~\cite{harper2015movielens} with 1 million records which consists of user's ratings of movies; 
\textbf{KuaiRand1K}\footnote{https://kuairand.com/} is a recent dataset for sequential short-video recommendation, and we use the 1K version~\cite{gao2022kuairand} which has irrelevant videos removed.
We preprocess all these datasets into a unified format where each record consists of (user features, user history, exposed items, user feedback, and timestamp) in sequential order.
The details of this process are summarized in Appendix \ref{apx: data_preprocess}, and the resulting dataset statistics is provided in Table \ref{tab: dataset}.

\begin{table}[t]
    \centering
    \begin{tabular}{c|cccc}
        \hline
        Dataset & $|\mathcal{U}|$ & $|\mathcal{I}|$ & \#record & $k$ \\
        \hline
        RL4RS & - & 283 & 781,367 & 9 \\
        MovieLens-1M & 6400 & 3706 & 1,000,208 & 10 \\
        KuaiRand & 986 & 11,643 & 969,831 & 10 \\
        \hline
    \end{tabular}
    \caption{Dataset Summary. $k$ represents the size of the recommendation list (i.e. the effect-action size). RL4RS dataset provides user profile features instead of user ID so it does not have a count for the user set.}
    \label{tab: dataset}
    \vspace{-10pt}
\end{table}

\subsubsection{Online Environment Simulator}\label{sec: simulator}
To simulate the online user interaction we train a user response model $\Psi: \mathcal{S}\times\mathcal{A} \rightarrow \mathbb{R}^k$ for each of the dataset.
The user state is derived based on the static user features and the dynamic history interactions.
$\Psi$ outputs the probabilities of the user's positive feedback on each item in recommended $a_t$, and the final response $y_t\in\{0,1\}^k$ (e.g. click) is uniformly sampled according to the probabilities.
We design the reward $r(s_t,a_t)\in[-0.2,1.0]$ as the average of the item-wise reward.
We provide details of the environment in Appendix \ref{apx: environment}.

\subsubsection{Models and Baselines}\label{sec: baselines}
We use SASRec~\cite{kang2018sasrec} as our backbone actor as described in section \ref{sec: method_state_encoder}, it consists of a Transformer-based state encoder and hyper-action generator, and the dot-product scorer.
We also implemented the following RL baselines using our proposed HAC framework to better showcase how our HAC generalizes existing methods:
\begin{itemize}
    \item \textbf{Online SL}: the SASRec actor directly learns from immediate user feedback instead of the long-term commutative reward.
    \item \textbf{A2C}: the synchronized version of the A3C~\cite{mnih2016asynchronous} that applies the policy gradient on the effect-action space.
    \item \textbf{DDPG}: a Deep DPG framework using the hyper-action space for both the actors and critics~\cite{timothy2016ddpg}. This method is equivalent to our HAC model without supervision.
    \item \textbf{TD3}: improve the DDPG with double Q-learning so the training of critic becomes more stable~\cite{fujimoto2018addressing}.
    \item \textbf{DDPG-RA}: the DDPG framework with the action representation learning as in~\cite{chandak2019learning}. This method is closest to our work and it regulates the effect-actions while our HAC model aligns the hyper-actions.
\end{itemize}
To better compare RL and SL methods, we also include the \textbf{Offline SL} that optimizes the policy using Eq.\eqref{eq: supervise_loss} based on the offline data instead of the online environment.
The model architectures and specifications of these models are provided in Appendix \ref{apx: model_specification}.

\begin{table*}[h!t]
    \centering
    \begin{tabular}{c|cccccc}
    \toprule
    \multirow{2}*{Model} & \multicolumn{2}{c}{RL4RS} &\multicolumn{2}{c}{ML1M} & \multicolumn{2}{c}{KuaiRand} \\
    & Total Reward & Depth & Total Reward & Depth & Total Reward & Depth \\
    \midrule
    Offline SL & 6.721 & 8.163 & 18.559 & 18.717 & \underline{14.394} & \underline{14.982} \\
    \midrule
    Online SL & \underline{9.502} & \underline{10.571} & \underline{18.629} & \underline{18.780} & 13.456 & 14.147 \\
    A2C & 7.789 & 9.140 & 16.158 & 16.556 & 12.460	& 13.250 \\
    DDPG & 8.337 & 9.588 & 17.205 & 17.508 & 11.394 & 12.313 \\
    TD3 & 8.553 & 9.791 & 17.545 & 17.814 & 11.777 & 12.664 \\
    PG-RA & 8.561 & 9.728 & 18.466 & 18.633 & 10.859 & 11.814 \\
    HAC & \textbf{10.059} & \textbf{11.102} & \textbf{18.863} & \textbf{18.988} & \textbf{14.789} & \textbf{15.335} \\
    \bottomrule
    \end{tabular}
    \caption{Online Performance. The best performances in bold and second best in Underline}
    \label{tab: main_results}
    \vspace{-10pt}
\end{table*}

\subsubsection{Evaluation}

For all datasets, we split them into the first 80\% for training and the last 20\% for evaluation according to record timestamps.
We then pretrain the online environment on the training set, and pretrain another online environment on the entire dataset for later evaluation.
We train our recommendation policies in the first environment and evaluate them in the second.
During training, we set the discount of reward as $\gamma=0.9$ and limit the interaction depth to $\leq 20$ for all experiments.
We find that most RL-based methods converge and stabilize within 50,000 iterations.
For long-term evaluation metrics, we consider the \textbf{Total Reward} that represents the summation of the rewards in an entire user session and the \textbf{Depth} represents how many interactions the user, and each session is observed using the simulated online environment that interacts with the learned policies.
And for both metrics, a higher value means better performance.
To evaluate the stability of the learned policies, we include a \textbf{Reward Variance} metric that estimates how inconsistent a policy would deal with different user states, so a lower value indicates a more stable policy.
Note that this metric describes the variance across states not the variance across random seeds.
In each experiment, we evaluate all aforementioned metrics and report the average values across different user sessions.

\subsection{Effectiveness}\label{sec: experiment_effectiveness}

For all tasks, the goal of the recommender system is to maximize the long-term satisfaction represented by total reward and average depth.
For each model, we grid-search the hyper-parameters and pick the setting with the best results to report in Table \ref{tab: main_results}.

\textbf{Main result:} We can see that the proposed HAC framework consistently achieves the best performances across all datasets on both long-term metrics: 6\% improvement on Rl4RS, 1\% on ML1M, and 3\% on KuaiRand over the best baselines.
This indicates the expressiveness of the proposed hyper-actions and the effectiveness of the learning method.
Note that all other RL methods can only achieve better results than offline supervised learning in the RL4RS task, but become worse in ML1M and KuaiRand with larger effect-action spaces.
This indicates that our HAC framework can better capture the patterns for larger action spaces.

\textbf{RL-baselines:} Among RL solutions, A2C always has the worst performance and appears to be the most unstable learning framework, but the gap between A2C and DDPG tends to be smaller in datasets (ML1M and KuaiRand) with larger action spaces and A2C even achieves better performances than DDPG in KuaiRand with the largest action space.
Since A2C directly optimizes the effect-action and DDPG uses the hyper-action, this reduced gap may indicate that it may become harder to learn consistent and accurate hyper-action representations in larger effect-action spaces.
This may also proves that ensuring consistency between the two action spaces is critical to achieving effective RL.
TD3 slightly improves the performance over DDPG but still behave in a similar way.

\textbf{Action space regularization: } In addition to our method, The DDPG-RA method also addresses the alignment of action spaces and has the closest behavior to our method.
Differently, it does not regulate the hyper-actions has HAC, instead, it aligns the effect-action space that is not directly used in guiding the actor.
Additionally, DDPG-RA uses the hyper-action rather than the effect-action when learning critics, so it achieves better results than A2C and DDPG, but does not surpass our method or even supervised methods.

\subsection{Learning HAC}\label{sec: experiment_learning}

\begin{figure}[t]
    \centering
    \includegraphics[width=\linewidth]{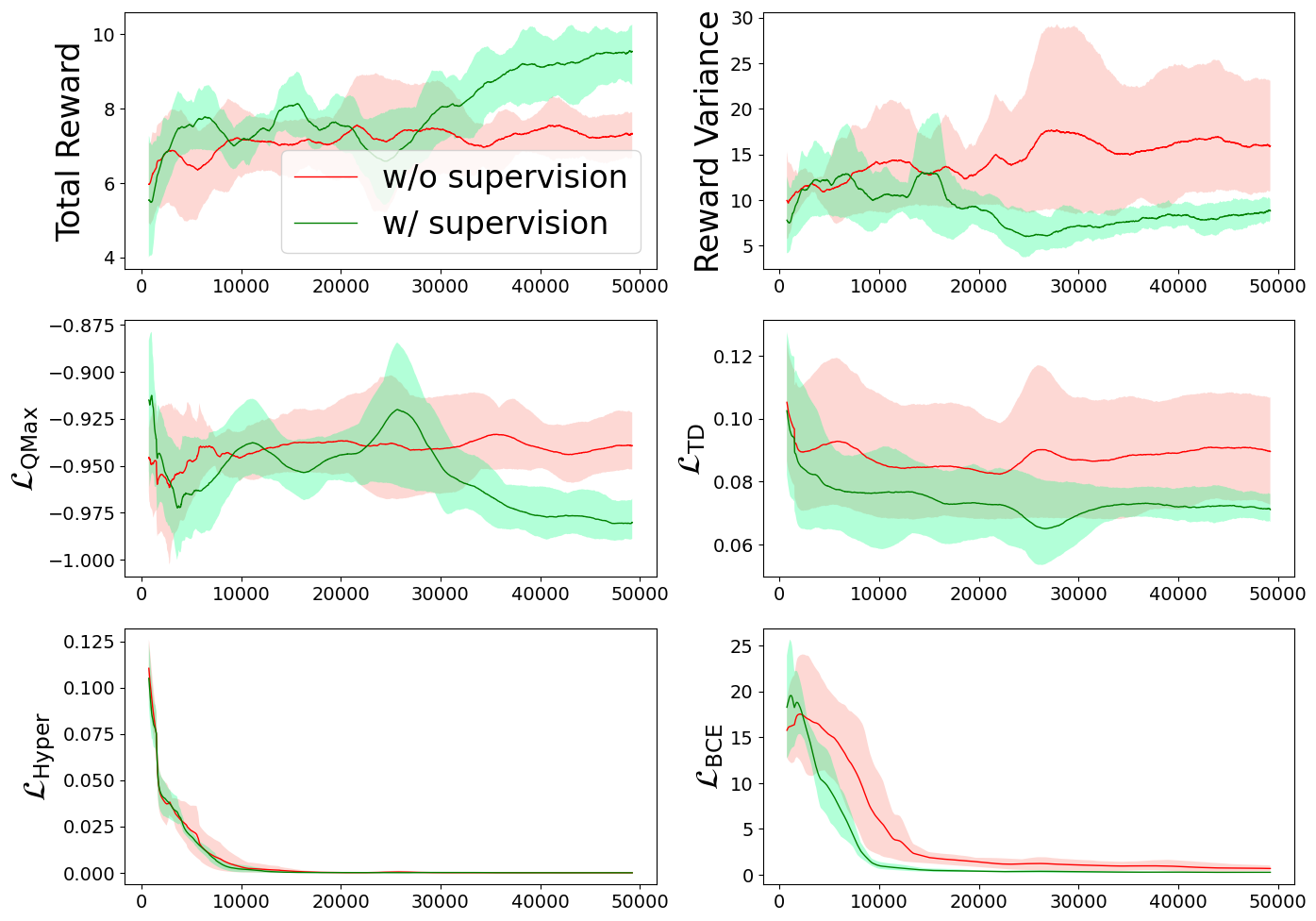}
    \caption{Training curves of HAC without supervision on RL4RS. X-axis corresponds to the number of iteration. The four losses are presented in log scales.}
    \label{fig: learning_curves_supervision}
    \vspace{-10pt}
\end{figure}

\begin{figure}[t]
    \centering
    \includegraphics[width=0.8\linewidth]{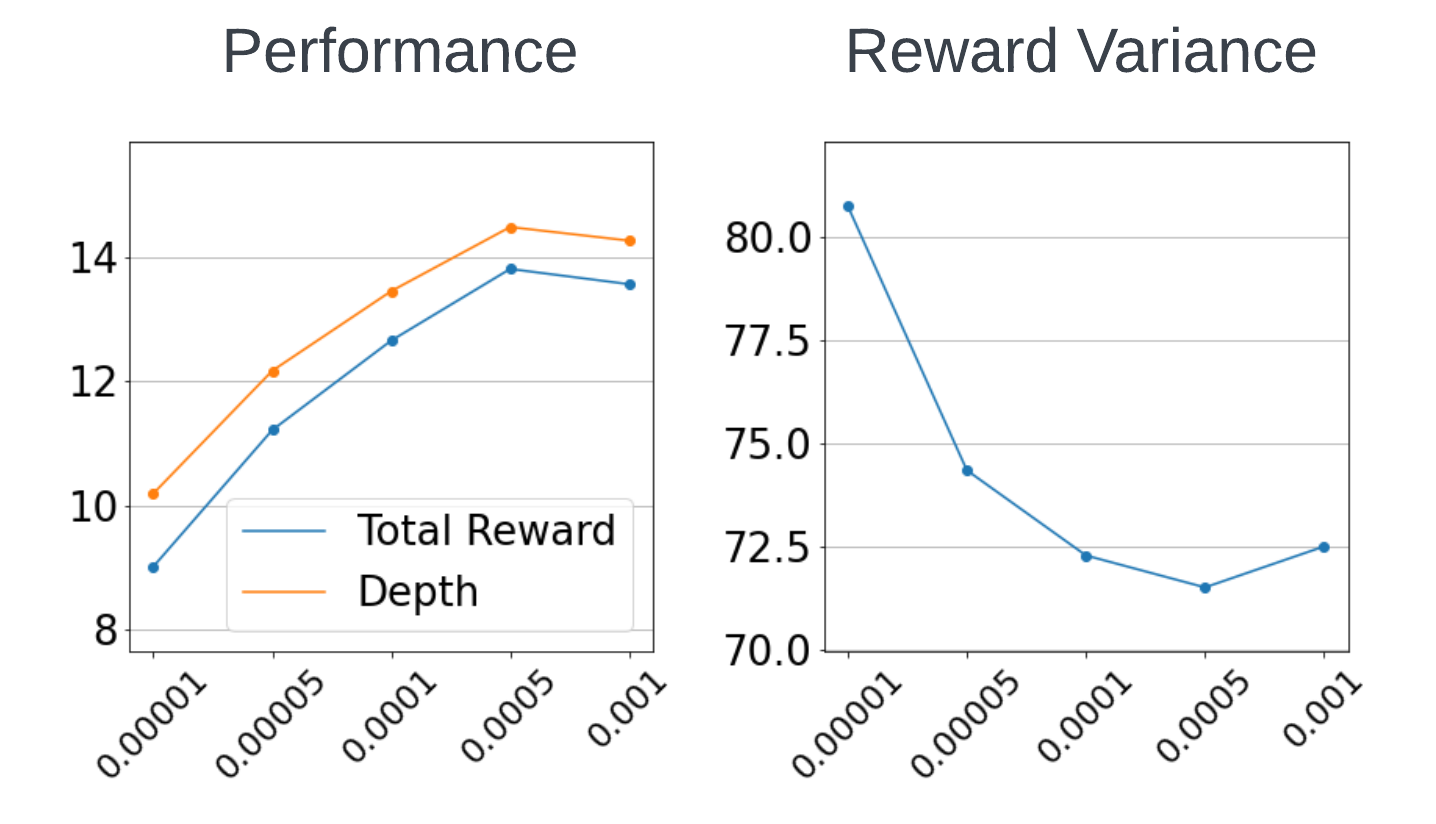}
    \caption{Effect of supervision loss on KuaiRand. X-axis represents the learning rate of Eq.\eqref{eq: supervise_loss}.}
    \label{fig: supervision_learning_rate}
    \vspace{-10pt}
\end{figure}

\textbf{Supervision and stable learning:} To further illustrate the training behavior of the HAC model, we plot the learning curves of HAC in Figure \ref{fig: learning_curves_supervision} where we compare a ``HAC w/o supervision'' that has no gradient for the supervised loss Eq.\eqref{eq: supervise_loss}.
All methods saturate on the performance metric as shown in the penal ``Total Reward'' and can successfully reduce all four loss functions mentioned in section \ref{sec: method_learning}.
Note that ``HAC w/o supervision'' can still successfully reduce the BCE loss on each item, indicating the effectiveness of RL based on the reward that aggregates the item-wise signals.
We can also see that including the supervision would help boost the model performance and reduce the actor loss $\mathcal{L}_\mathrm{QMax}$ that helps explore better actions.
Note that HAC has a lower critic loss $\mathcal{L}_\mathrm{TD}$ than ``HAC w/o supervision'', which indicates a more stable learning process.
We can also verify this by observing the variance of the total reward across users since higher reward variance indicates that the learned policy is less capable of providing good actions under different user states.
As shown in Figure \ref{fig: supervision_learning_rate} for the KuaiRand environment, increasing the importance of supervised loss would help improve the recommendation performance and reduce the variance.
Yet, assigning the supervision module with a learning rate (0.001 in KuaiRand) that is too large may over-exploit the user's intention and harm the performance.
We observe similar patterns in the other two datasets and provide the results in Appendix \ref{apx: experiments}.

\begin{figure}[t]
    \centering
    \includegraphics[width=\linewidth]{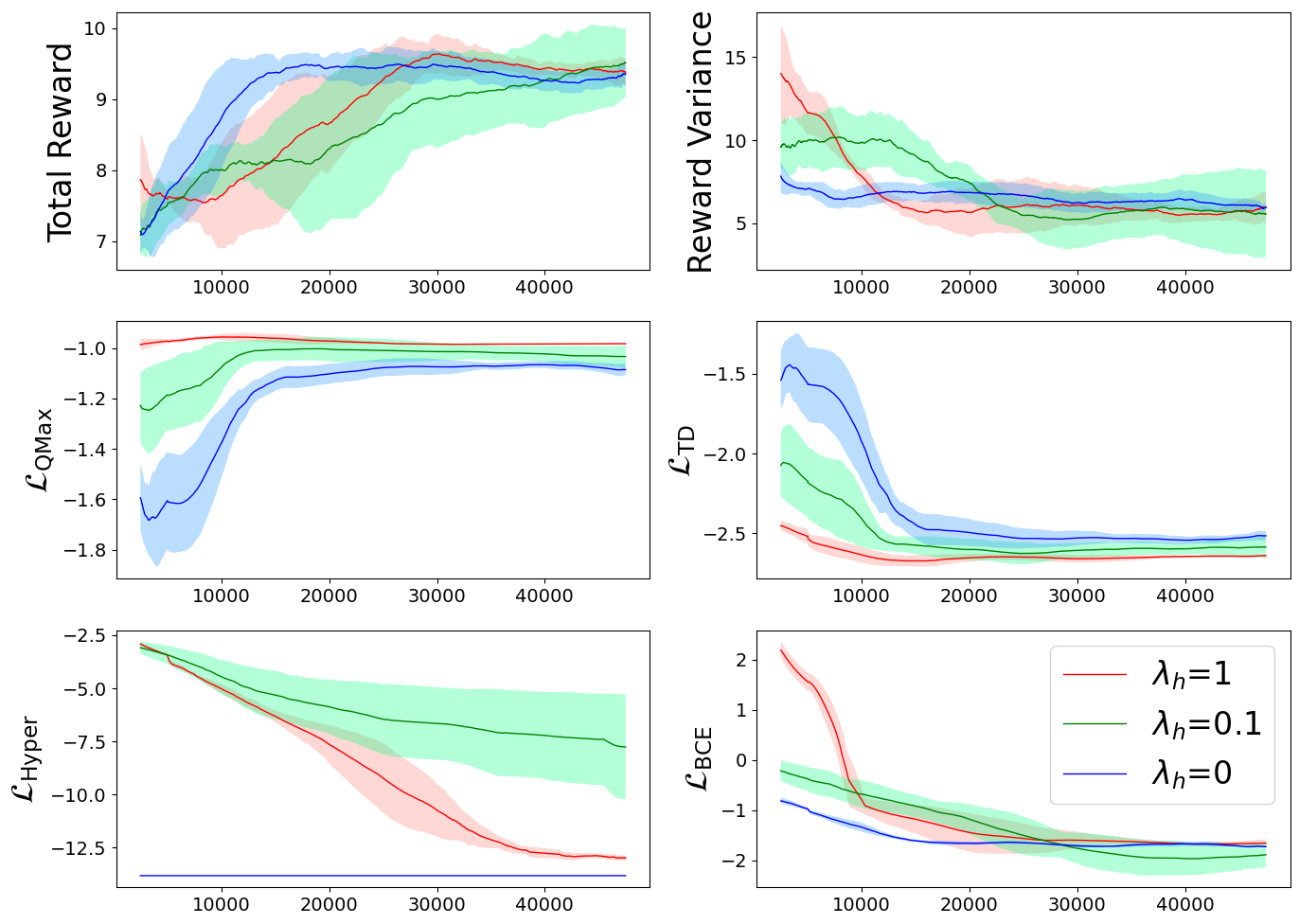}
    \caption{Training curves of HAC on RL4RS. The reward variance correspond to the variance of total reward across different users. The four losses are presented in log scales. $\lambda_h$ represents the magnitude of hyper-action alignment. $\lambda_h=0$ does not include this alignment loss so has $\mathcal{L}_\mathrm{Hyper}=0$.}
    \label{fig: learning_curves_alignment}
    \vspace{-10pt}
\end{figure}

\begin{figure}[t]
    \centering
    \includegraphics[width=0.8\linewidth]{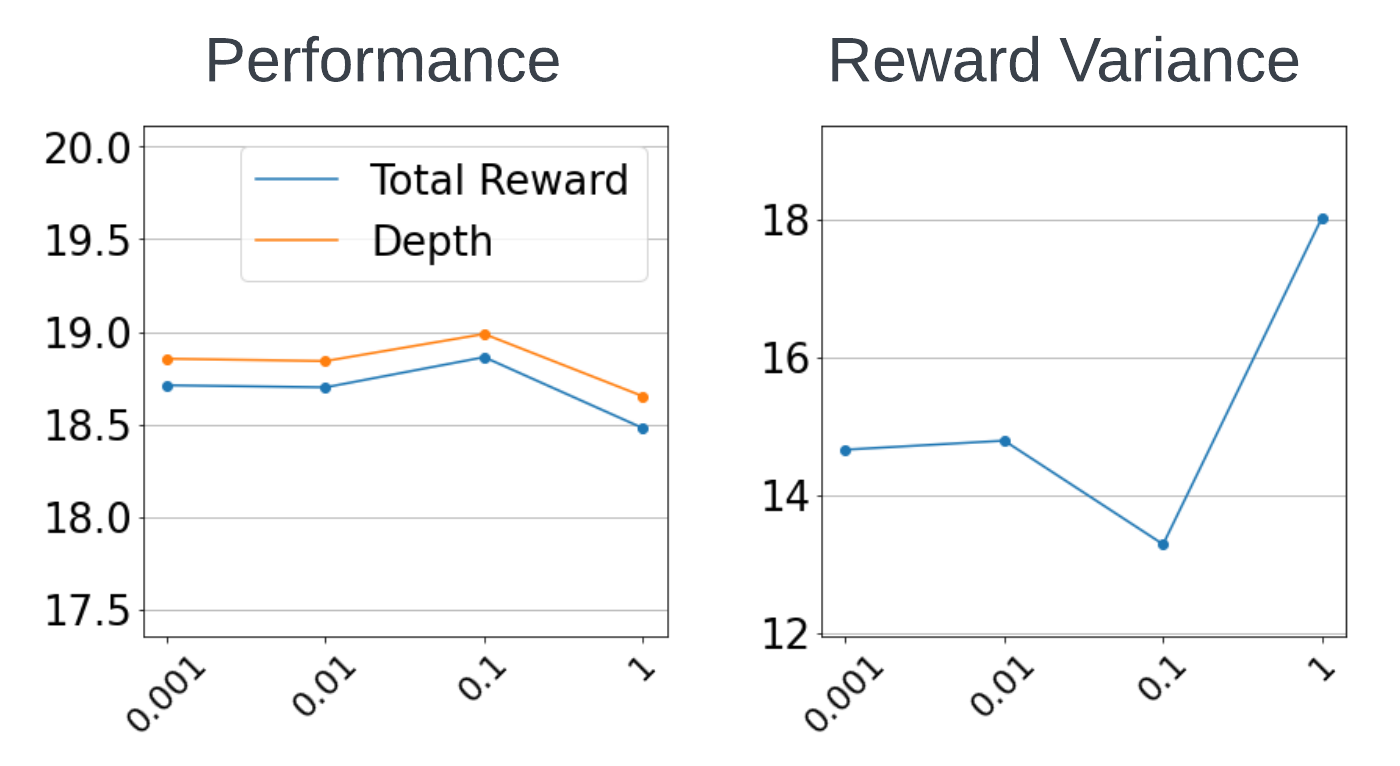}
    \caption{Effect of hyper-action alignment on ML1M. X-axis represents the magnitude of Eq.\eqref{eq: hyper_actor_loss}, i.e. $\lambda_h$ in Figure \ref{fig: learning_curves_alignment}.}
    \label{fig: hyper_learning_rate}
    \vspace{-10pt}
\end{figure}

\textbf{Hyper-action regulation:} We conduct the same ablation experiments when comparing different magnitudes of hyper-action alignment for loss Eq.\eqref{eq: hyper_actor_loss} and plot the learning curves of HAC in Figure \ref{fig: learning_curves_alignment}
In general, including the hyper-action alignment ($\lambda_h=0.1$ and $\lambda_h=1$) would cause the HAC framework to learn slower than the HAC that only uses supervision ($\lambda_h=0$) in terms of the convergence of total reward and reward variance.
In contrast, the more consistent action space helps the model to learn and explore better action policies.
Besides, increasing the importance of this alignment module results in worse $\mathcal{L}_\mathrm{QMax}$ and better $\mathcal{L}_\mathrm{TD}$, indicating that critic is more accurate in capturing the quality of actions.
Note that $\lambda_h=1$ is more stable than $\lambda_h=0.1$ but may be less effective in exploration.
To better verify this We illustrate the evaluation result in Figure \ref{fig: hyper_learning_rate} where the results exhibit a best point $\lambda_h=0.1$ where the recommendation is higher and more stable in reward.

\subsection{Ablation Study}\label{sec: experiment_ablation}

\begin{figure}[t]
    \centering
    \includegraphics[width=\linewidth]{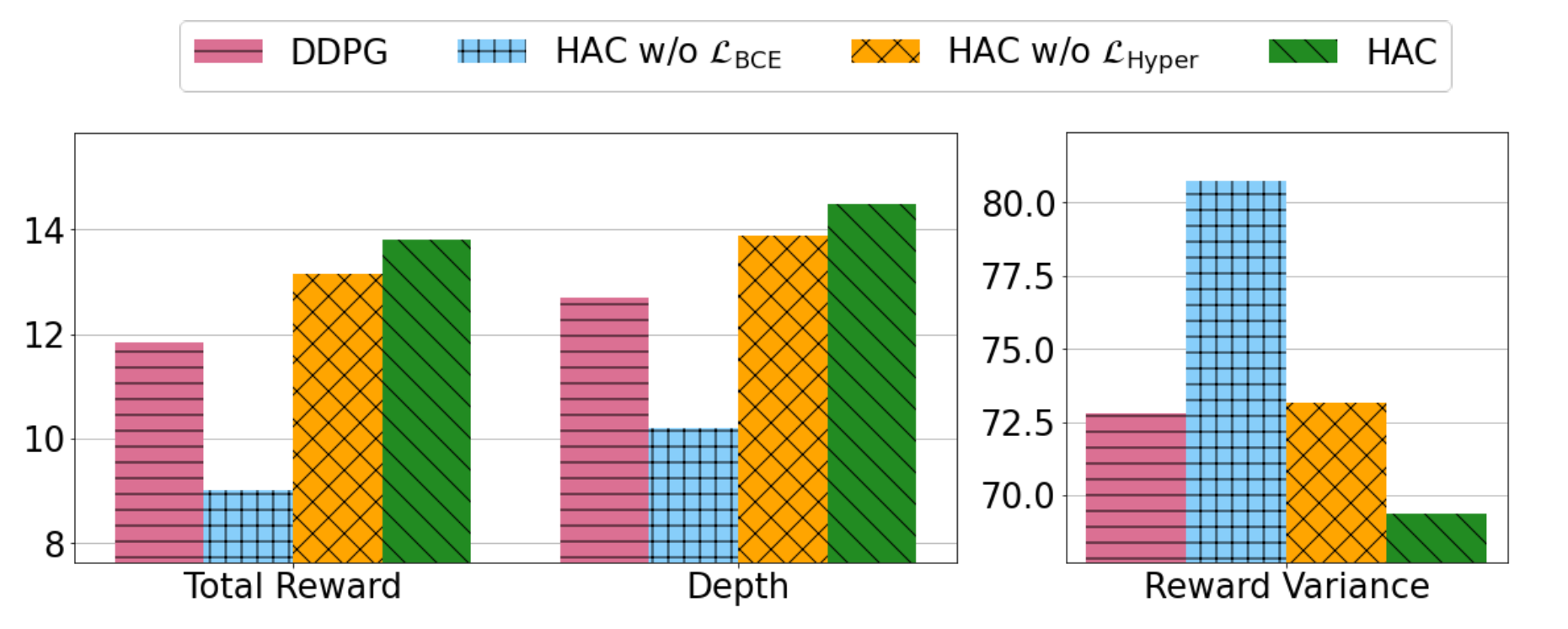}
    \caption{Effect of model components on ML1M.}
    \label{fig: ablation}
    \vspace{-10pt}
\end{figure}

\textbf{Model components:} To better investigate how different learning modules in our framework work, we compare several alternatives of our method:
1) DDPG: HAC without supervision and action alignment, and it uses hyper-action space for both actor learning and critic learning;
2) HAC w/o $\mathcal{L}_\mathrm{BCE}$: excluding the supervision of HAC;
3) HAC w/o $\mathcal{L}_\mathrm{Hyper}$: excluding the hyper-action alignment module.
We summarize the results in Figure \ref{fig: ablation} for ML1M dataset.
We can see that excluding either the supervision or the alignment module would reduce the performance and increase the reward variance.
This indicates that both modules help improve the model's accuracy and learning stability.
Similar results are also observed in other datasets as augmented in Appendix \ref{apx: experiments}.
Note that DDPG achieves relatively the same performance as HAC w/o $\mathcal{L}_\mathrm{BCE}$, this indicates that using separating action spaces for actor learning and critic learning as in HAC may reduce the performance and simply including an action alignment module would not fill the gap.
In this sense, HAC needs both hyper-action alignment and supervision.
This also means that the inconsistency between the two action spaces is smaller than the inconsistency between the aggregated reward and item-wise user response signals.

\begin{figure}[t]
    \centering
    \includegraphics[width=0.8\linewidth]{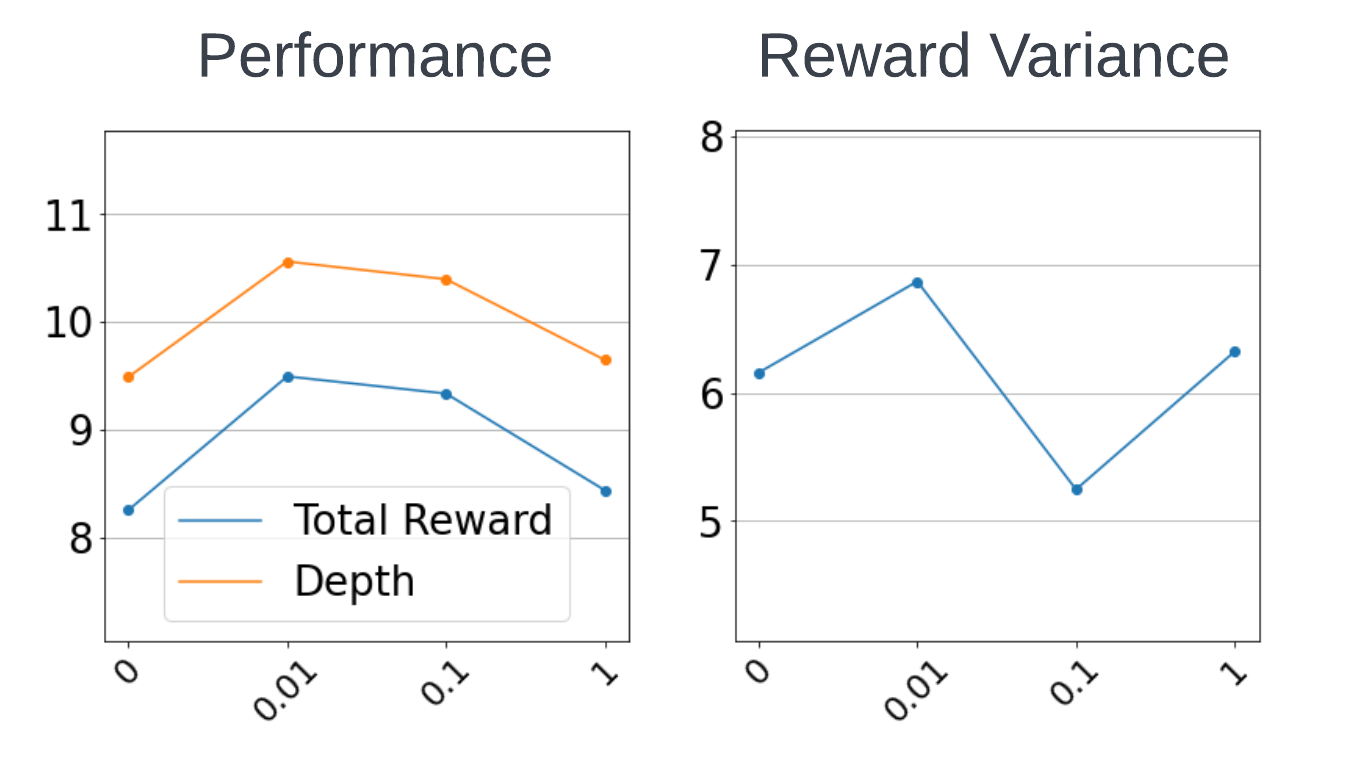}
    \vspace{-10pt}
    \caption{Effect of HAC's hyper-action exploration magnitude on RL4RS dataset. X-axis correspond to the Gaussian noise variance for hyper-actions in HAC model.}
    \label{fig: noise_rl4rs}
    \vspace{-10pt}
\end{figure}

\begin{figure}[t]
    \centering
    \includegraphics[width=0.8\linewidth]{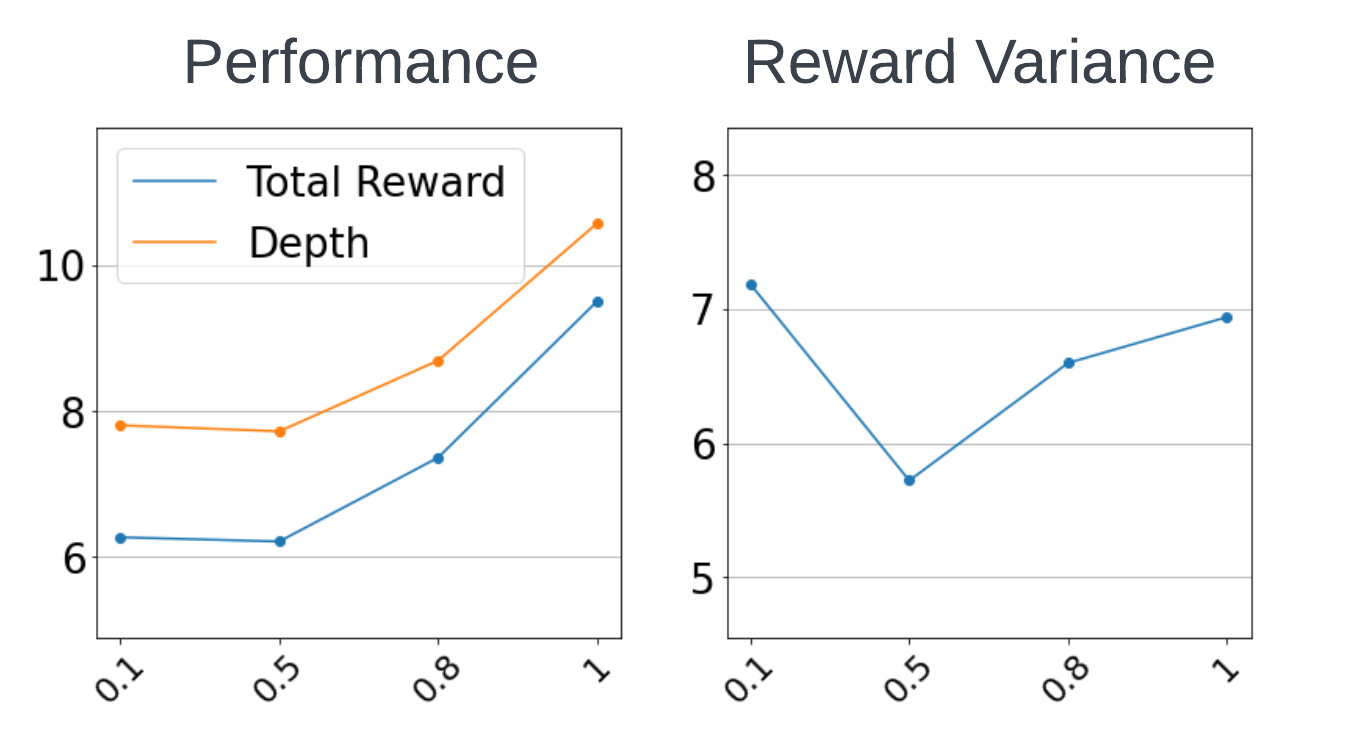}
    \vspace{-10pt}
    \caption{HAC's effect-action exploration magnitude on RL4RS dataset. X-axis correspond to the rate of greedy top-k selection rather than categorical sampling for effect-actions in HAC model, and we set hyper-action noise to 0.}
    \label{fig: noise_rl4rs_topk}
    \vspace{-10pt}
\end{figure}

\textbf{Exploration on different action spaces: } In terms of the exploration of HAC model, we can apply exploration on both the hyper-action space and the effect-action space.
To compare the effect of different magnitude of hyper-action exploration, we change the variance of the Gaussian noise during learning and fix all other hyperparameters of HAC, and present the results in Figure \ref{fig: noise_rl4rs}.
The comparison of recommendation performance shows an optimal point in the middle of the search space, indicating that one should carefully design the exploration so that the sampling variance is not too small or too large.
As we have discussed in section \ref{sec: method_exploration}, 
small variance may limit the exploration of new actions and large variance may induce unstable exploration that hardly converges.
And empirically, we find that sampling on effect-actions is less effective than exploration on hyper-actions.
As the example in Figure \ref{fig: noise_rl4rs_topk}, applying top-$k$ greedy selection achieves the best result and adding categorical sampling would make the learned policy sub-optimal.

\section{Related Work}\label{sec: related_work}

\subsection{Sequential Recommendation and Session-based Recommendation}\label{sec: related_work_seqrec}

The session-based recommendation (SBR) problem is closely related to the sequential recommendation (SR) task~\cite{wang2021survey}.
An SR task aims to learn a policy that can infer the next recommendation (item or list) based on the given user's historical interactions.
In comparison, the SBR considers the existence of the beginning and termination of an interaction session.
We see our setting as somewhere intersects the two notions: by setting the list size to 1, it would be almost identical to SR except for the reward setup; yet, our goal is to optimize the entire future reward of the user session, which is closer to the next-partial SBR as defined in ~\cite{wang2021survey}.
Under both problem settings, the most adopted solution uses the Markov Chain assumption to model the dynamic transitions of the user-system interactions.
The main challenge of this line of work is how to construct a representative user state based on the long-term history.
Early solutions to the recommendation problem adopted the collaborative filtering techniques~\cite{rendle2010factorizing,cheng2016widedeep,liang2018variational,koren2022advances} and later approaches embrace deep neural networks like Recurrent Network~\cite{jannach2017gru4rec}, Convolution Network~\cite{tang2018personalized}, Memory Network~\cite{chen2018sequential}, Self-Attention~\cite{kang2018sasrec, sun2019bert4rec}, GCN~\cite{wu2019session,chang2021sequential}, Machine Reasoning \cite{shi2020neural,chen2021neural,ji2023counterfactual} and Foundation Models \cite{geng2022recommendation} to improve the model expressiveness, so that it can better capture the abundant and complex information from user/item features and interaction histories.
The key insight behind all these methods is to accurately encode the long-range histories, but this paradigm does not optimize the long-term user rewards.

\subsection{\mbox{Reinforcement Learning in Recommendation}}\label{sec: related_work_rl4rs}

The RL-based RS ~\cite{shani2005mdp,sutton2018reinforcement,afsar2021reinforcement} also follows the MDP formulation and it emphasizes the importance of optimizing the cumulative reward that represents long-term user satisfaction.
In the simplest setting, tabular-based methods ~\cite{mahmood2007learning} are used to optimize an evaluation table but only work for a fixed set of state-action pairs. 
Then value-based methods ~\cite{taghipour2007usage,zhao2018recommendations,zhao2021dear,pei2019value} and policy gradient methods ~\cite{chen2019top,chen2019large,ge2021towards,xian2019reinforcement,ge2022toward,li2022autolossgen} are proposed to learn to evaluate and optimize the action policy based on the sampled long-term reward.
The actor-critic paradigm ~\cite{zhao2018deep,xin2020self,zhao2020whole} integrates these two methods by simultaneously learning an action evaluator and an action generator.
The main challenges of RL-based RS consist of the large combinatorial state/actions space ~\cite{dulac2015deep,ie2019slateq,liu2020state}, regulating the unstable learning behavior ~\cite{bai2019model,chen2021user}, and finding optimal reward function for heterogeneous user behaviors~\cite{chen2021generative, cai2022constrained}.
Our work focus on the action space representation learning and the stability of RL.
And we consider PG-RA~\cite{chandak2019learning} as the closest work that also emphasizes the learning of latent action representations.
As we have mentioned in section \ref{sec: experiment_setup}, PG-RA aims to transfer knowledge through a shared scoring function and applies action alignment on the effect-action space, which is not well suited for the latent-factor decomposition for users and items.
We have empirically verified this inferior performance in \ref{sec: experiment_effectiveness}.
Additionally, we have illustrated the effectiveness of our method through evaluation on different simulated environments, but we remind readers that there is still a chance that the actual online environment is a more complex and dynamic mechanism.
In this sense, there are works focusing on building a more realistic online user environment for RS ~\cite{ie2019recsim,zhao2019deep,zhao2021usersim} which could complement our work in practice.

\section{Conclusion}\label{sec: conclusion}

We proposed a practical actor-critic learning framework that optimizes a recommendation policy in both the discrete effect-action space as well as the vectorized hyper-action space.
Empirical results verified the effectiveness and stability of our proposed method in online environments built on public datasets.
The main purpose of our design is the generalization ability that can accommodate various existing solution frameworks including Actor-Critic, DDPG, as well as supervised learning.
And we consider the notion of hyper-action as a preliminary attempt on applying more sophisticated scoring functions than a simple dot product.

\balance

\begin{acks}
This research was partially supported by APRC---CityU New Research Initiatives (No.9610565, Start-up Grant for New Faculty of City University of Hong Kong), SIRG---CityU Strategic Interdisciplinary Research Grant (No.7020046, No.7020074), and CityU---HKIDS Early Career Research Grant (No.9360163)
\end{acks}

\bibliographystyle{ACM-Reference-Format}
\bibliography{main}

\newpage

\appendix

\section{Experiment Setup}

\subsection{Dataset Processing}\label{apx: data_preprocess}

For the ML1M dataset, we treat movies with user ratings higher than 3 as positive samples (representing a ``like'') and others as negative samples.
For the KuaiRand\cite{gao2022kuairand} dataset, we first apply 50-core filtering where videos with less than 50 occurrences are removed.
Videos with watch time ratio larger than 0.8 are regarded as positive samples (representing a ``valid view'' or a ``comsumption'').
For both ML1M and KuaiRand datasets, we segment each user session into a sequence of item lists of length 10 chronologically.
For each segmented list, only the positive samples before this list are viewed as the historical behavior.
So each record in the final dataset consists of user id, history behavior sequence, item lists and label lists, which is in the same format as the RL4RS dataset.

\subsection{Online Environment}\label{apx: environment}

The environment is built based on a user response model that consists of an item embedding layer, an attention-based history encoder, and a scoring module for item in actions. 
The embedding layer is responsible for converting sparse features into dense features and it is shared by the later history encoder and action item scoring module.
The history encoder layer consists of two multi-head-attention and output a user embedding. 
The first network encodes the sequence information in the history items, and the second layer encodes both the sequence information from previous layer and the user information. 
The final scoring module calculate the dot product between the target/action item embedding and the user embedding, and we adopt binary cross-entropy loss during training.
As we have mentioned in section \ref{sec: experiment_setup}, the user's feedback is sampled according to these final scores.
After receiving a recommendation list and responding with the user's feedback, the positive items are used to update the user's history while the static features stay the same.
We design the reward $r(s_t,a_t)$ as the average of the item-wise reward, where a positive response gives a reward of 1 and a negative response gives a reward/cost of -0.2.

We also construct a temper-based user leave model to give the end-of-session signal $d$, and the user leaves the system when the system used up the user's patience so there is no remaining temper to engage another round of interaction.
The user leave model maintains a users budget of temper, and the budget decreases whenever a user received a recommendation list.
The magnitude of temper budget reduction is determined by the quality of recommendation, and worse recommendation induces quicker temper drops.
Additionally, we restrict the maximum session depths to 20 for all environments.

\subsection{Model Specifications}\label{apx: model_specification}

\begin{table*}[t]
    \small
    \centering
    \begin{tabular}{c|cccc}
        \hline
        Method & Actor Learning & Critic Learning & Supervised Learning & Action Alignment \\
        \hline
        SL & None & None & BCE & None \\
        A2C & Policy gradient on $P(a|s)$ with $V(s)$ & TD error on $V(s)$ & None & None \\
        DDPG & $\max Q(s,Z)$ with $P(Z|s)$ & TD error on $Q(s,Z)$ & None & None \\
        TD3 & $\max \min(Q_1(s,Z),Q_2(s,Z))$ with $P(Z|s)$ & TD error on $\min(Q_1(s,Z),Q_2(s,Z))$ & None & None \\
        DDPG-RA & $\max Q(s,Z)$ with $P(Z|s)$ & TD error on $Q(s,Z)$ & None & $-\mathbb{E}\Big[\ln(P(a_t|s_t,s_{t+1}))\Big]$ \\
        HAC & $\max Q(s,Z)$ with $P(Z|s)$ & TD error on $Q(s,a)$ & Optional & Eq.\eqref{eq: hyper_actor_loss} \\
        \hline
    \end{tabular}
    \caption{Method Component Comparison}
    \label{tab: existing_solution}
\end{table*}

In this part we present the architectures and hyperparameters of models.
Our code is implemented in PyTorch and we provide source code in \href{https://github.com/CharlieMat/Hyper-Actor-Critic-for-Recommendation}{https://github.com/CharlieMat/Hyper-Actor-Critic-for-Recommendation}.

\textbf{SASRec Backbone Policy: } This module is shared across all methods in Table \ref{tab: existing_solution}. For any given item embedding (appeared in the history sequence, the exposure, and the candidate item pool), an item encoder is used to map it to a 32-dimensional vector.
Then the encoded vectors of the user history is added with a trainable positional embedding and concatenated with the encoded user vector (by a 2-layer  with hidden dim 128).
The final input sequence is then input into a 2-layer transformer encoder with 4 heads and dropout rate 0.1. 
The last embedding in the output sequence is regarded as the encoded user state.
Finally, we use a linear mapping to generate the hyper-action from the user state.
And the final effect-action is selected based on the dot-product ranking score as described in section \ref{sec: method_scorer}.

\textbf{Critic: } the critic is a MLP with (encoded user state vector, hyper-action vector) as input, two hidden layers with dimension 256 and 64, and outputs a scalar representing the Q value.
Differently, in A2C, a value network is used which has the same hidden dimensions but only takes the user state as input.

\textbf{Offline Supervise: } we use the offline training set to optimize the policy through binary cross entropy loss Eq.\eqref{eq: supervise_loss} based on the observed exposure (a list of recommend items) and the corresponding user feedback.
We adopt batch size 64, and searched the learning rate in [0.0005, 0.0001, 0.00005, 0.00001] and L2 regularization with coefficient in [0.0001, 0.00005, 0.00001, 0.000005].
When evaluating the online performance during training, we consider apply mini-batch training for each interaction step.

\textbf{Online Supervise: } the online version of SL differs from the offline SL by directly using the sampled trajectories collected by online interactions with users rather than mini-batches from offline datasets.

\begin{figure}[t]
    \centering
    \includegraphics[width=\linewidth]{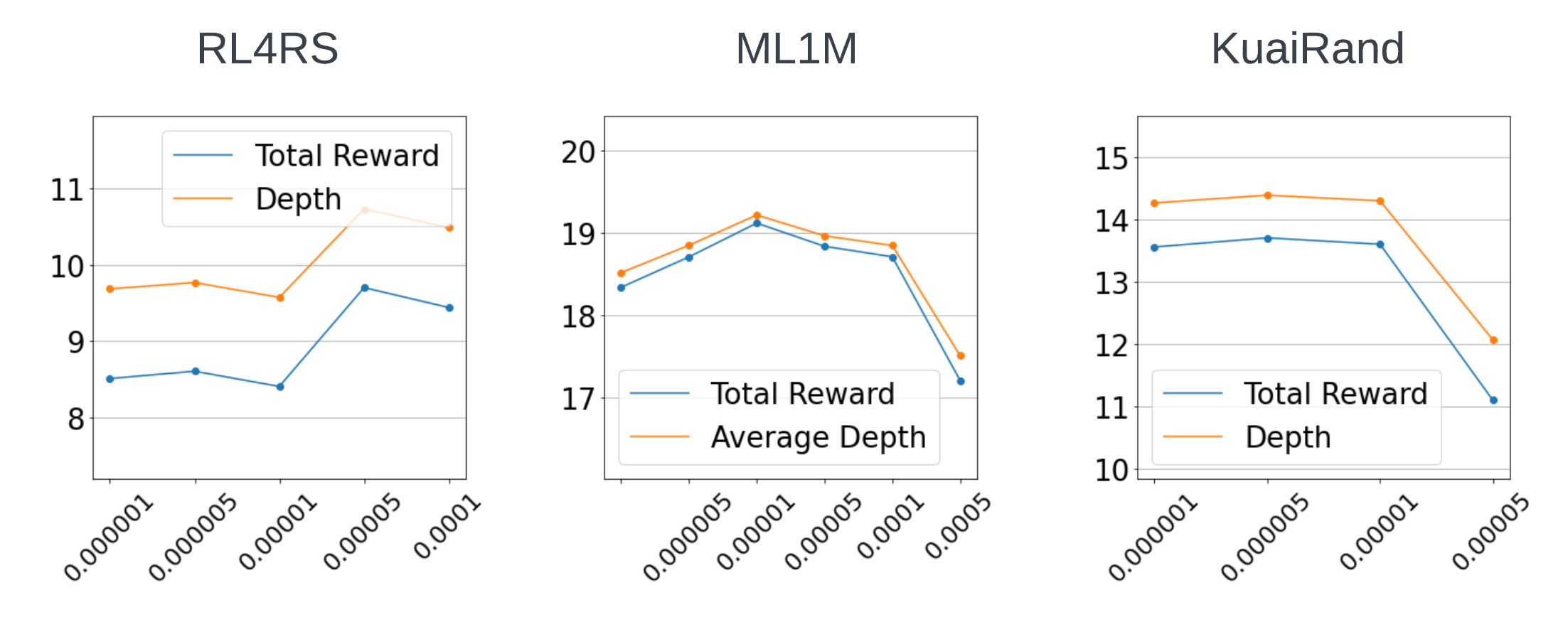}
    \caption{Controlling Actor Learning Rate of HAC, X-axis represents learning rate of $\mathcal{L}_\mathrm{QMax}$.}
    \label{fig: actor_learning_rate}
\end{figure}

\begin{figure}[t]
    \centering
    \includegraphics[width=0.8\linewidth]{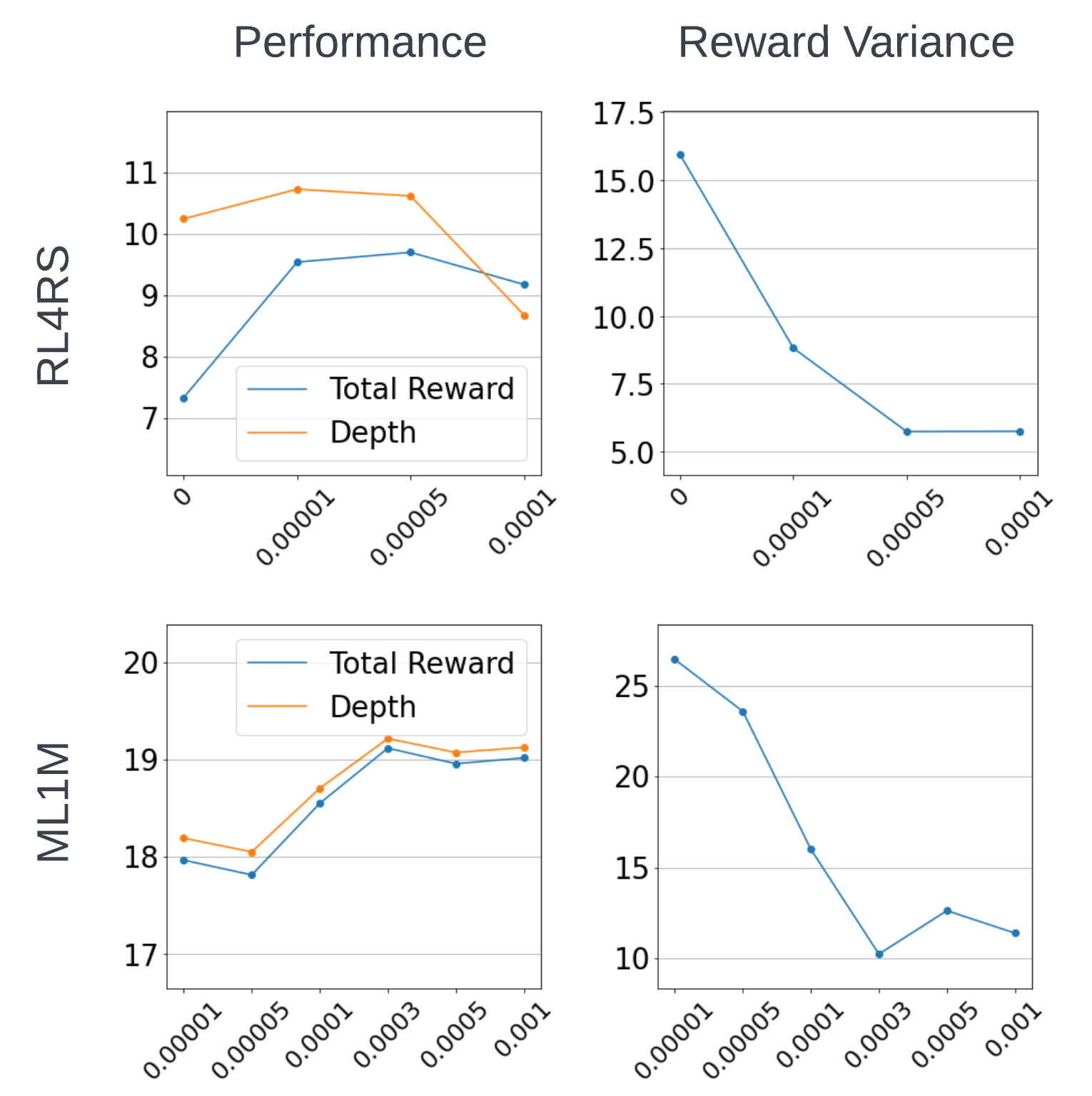}
    \caption{Supervision learning rate of Eq.\eqref{eq: supervise_loss} on ML1M and RL4RS datasets.}
    \label{fig: supervision_lr_other_datasets}
\end{figure}

\begin{figure}[t]
    \centering
    \includegraphics[width=0.8\linewidth]{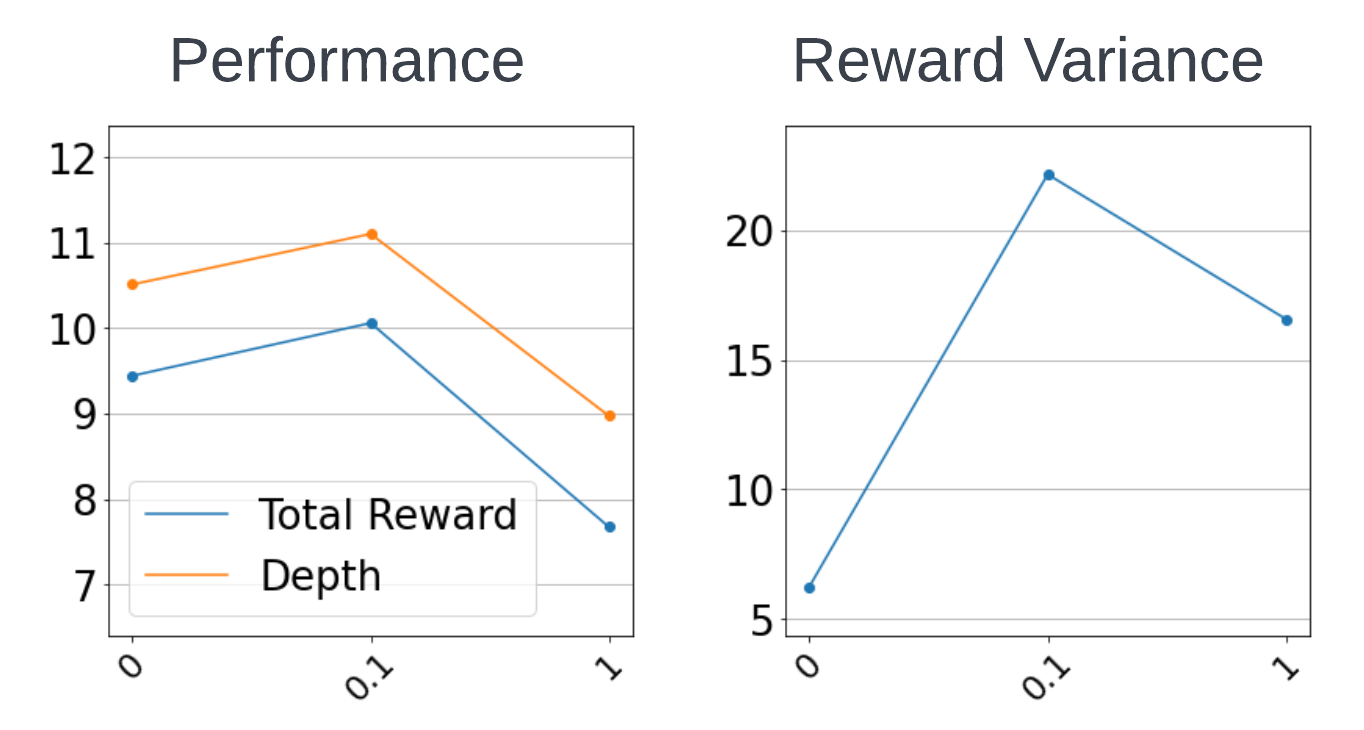}
    \caption{Effect of hyper-action alignment on RL4RS online environment. X-axis represents the magnitude of Eq.\eqref{eq: hyper_actor_loss}.}
    \label{fig: hyper_lr_rl4rs}
\end{figure}

For all the following RL-based methods, we use a replay buffer with size 100,000 and starts mini-batch training when the buffer has more than 2,000 samples.
Each sample in the buffer consists of (user static features, current history, next history, hyper-action embedding, effect-action the recommendation list, user response, immediate reward, and the ``done'' signal).
And we simultaneously run 32 user sessions/episodes so each interaction step will sample 32 records into the buffer.
The current user state combines (user static features, current history) and the next user state combines (user static features, current history).
The common search spaces of hyperparameters are [0.0005, 0.0001, 0.00005, 0.00001, 0.000005, 0.000001] for actor learning rate, [0.001, 0.0001, 0.00001] for critic learning rate, and variance $\sigma\in[0.1, 0.01, 0.001]$ for the Gausian hyper-action exploration.
In addition to the model differences mentioned in Table \ref{tab: existing_solution}, we summarize the experiment settings as the following:

\textbf{A2C: } this method combines the aforementioned SASRec backbone and the value (critic) network.
uses the policy gradient that is guided by the advantage score of the value.
The exploration of A2C does not explore on hyper-actions so has noise variance $\sigma=0$, and the action exploration is engaged based on categorical sampling on the effect-action probabilities.

\textbf{DDPG: } this model uses the Q-network for critic and adopt Gausian sampling on the hyper-action space.

\textbf{DDPG-RA: } this model is the DDPG model with an additional effect-action alignment module as illustrated in Table \ref{tab: existing_solution}.
For the alignment module, we search the learning rate in [0.00005, 0.00001, 0.000005, 0.000001].

\textbf{HAC: } it uses the effect-action space for critic learning and hyper-action space for actor learning.
As a result, we can apply exploration on both action spaces.
Compared to DDPG and supervised learning, this method has an additional action alignment module where we search the magnitude in $\lambda_h\in[0.001, 0.01, 0.1, 1]$ as described in section \ref{sec: experiment_learning}.

\section{Auxiliary Experimental Results}\label{apx: experiments}

\begin{figure}[t]
    \centering
    \includegraphics[width=\linewidth]{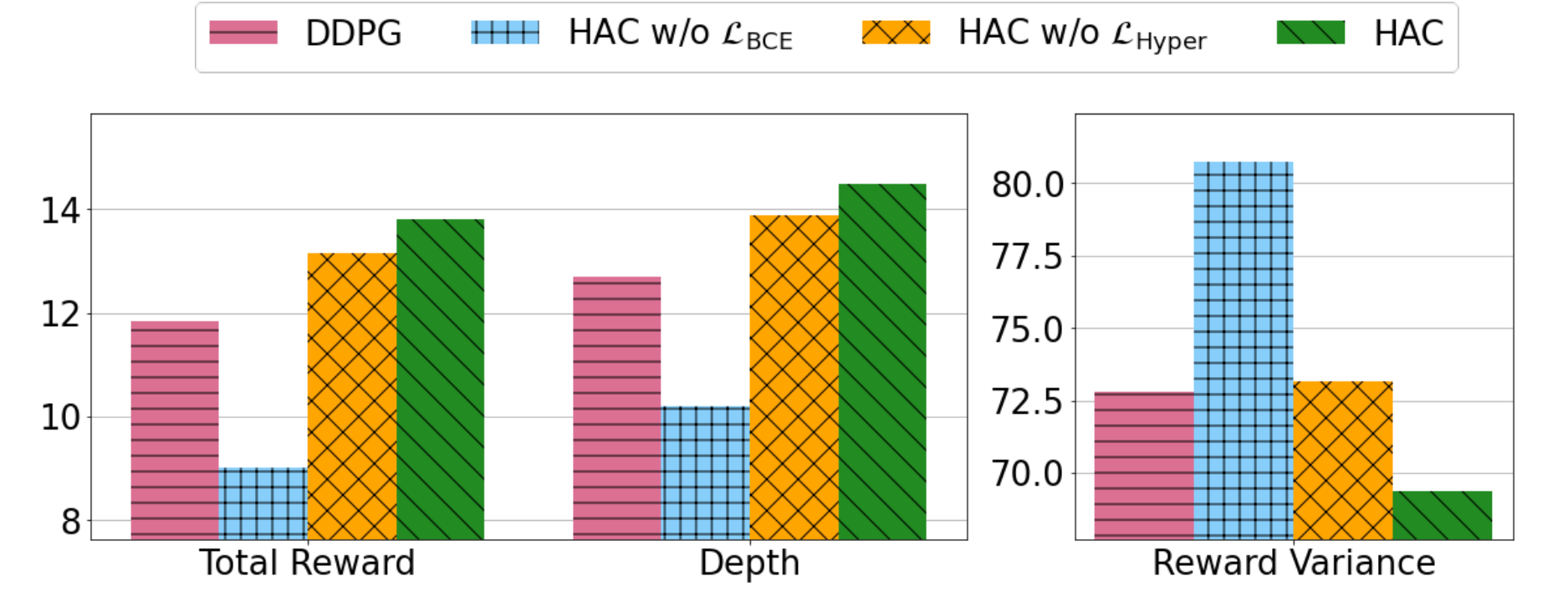}
    \caption{HAC model ablation results on KuaiRand online environment}
    \label{fig: ablation_kuairand}
\end{figure}

We present comparison results of actor learning rates of HAC in Figure \ref{fig: actor_learning_rate}.
We present comparison results of different supervision learning rates of HAC in Figure \ref{fig: supervision_lr_other_datasets} as the augmentation of experimental results in section \ref{sec: experiment_learning}.
We present comparison of different model ablation of HAC in Figure \ref{fig: ablation_kuairand} as the augmentation of experimental results on ablation study in section \ref{sec: experiment_ablation}.
All results are consistent with the results in the main experimental results which further verifies our claims.

\end{document}